\documentclass[preprint, pra, aps, superscriptaddress, tightenlines]{revtex4-1}

\usepackage{inputenc}
\usepackage{graphicx}
\usepackage{hyperref}
\usepackage{textcomp}
\usepackage{amssymb, amsmath}

\begin{document}

\title{Tunable Squeezing Using Coupled Ring Resonators on a Silicon Nitride Chip}

\author{Avik Dutt}
\email{ad654@cornell.edu}
\author{Steven Miller}
\author{Kevin Luke}
\author{Jaime Cardenas}
\affiliation{School of Electrical and Computer Engineering, Cornell University, Ithaca, NY 14853, USA}

\author{Alexander L. Gaeta}
\affiliation{School of Applied and Engineering Physics, Cornell University, Ithaca, NY 14853, USA}

\author{Paulo Nussenzveig}
\affiliation{Instituto de Física, Universidade de São Paulo, P.O. Box 66318, 05315-970 São Paulo, Brazil}

\author{Michal Lipson}
\affiliation{School of Electrical and Computer Engineering, Cornell University, Ithaca, NY 14853, USA}
\affiliation{Kavli Institute at Cornell for Nanoscale Science, Cornell University, Ithaca NY 14853, USA}


\begin{abstract}
We demonstrate continuous tuning of the squeezing level generated in a double-ring optical parametric oscillator by externally controlling the coupling condition using electrically controlled integrated microheaters. We accomplish this by utilizing the avoided crossing exhibited by a pair of coupled silicon nitride microring resonators. We directly detect a change in the squeezing level from 0.5 dB in the undercoupled regime to 2 dB in the overcoupled regime, which corresponds to a change in the generated on-chip squeezing factor from 0.9 dB to 3.9 dB. Such wide tunability in the squeezing level can be harnessed for on-chip quantum enhanced sensing protocols which require an optimal degree of squeezing.
\end{abstract}

\maketitle

Squeezed states of light \cite{lvovsky_squeezed_2014} are an important resource for a wide range of applications such as quantum-enhanced sensing \cite{taylor_biological_2013}, spectroscopy \cite{polzik_spectroscopy_1992, ribeiro_sub-shot-noise_1997} and metrology \cite{giovannetti_quantum-enhanced_2004, taylor_quantum_2014}, as well as for quantum information processing in the continuous variable regime \cite{braunstein_quantum_2005, obrien_photonic_2009}. Generation of squeezed light was first reported in the seminal experiments of Slusher \emph{et al.} using atomic vapors in an optical cavity \cite{slusher_observation_1985}. The workhorse for generating squeezed light for the past three decades has been the optical parametric oscillator (OPO), both below \cite{wu_generation_1986, wu_squeezed_1987} and above threshold \cite{heidmann_observation_1987, furst_quantum_2011}. An OPO typically consists of a second- or third-order nonlinear medium placed in a cavity \cite{giordmaine_tunable_1965}. The degree of squeezing generated by the OPO, defined as the quantum noise reduction below the shot noise level, depends on the ratio of the coupling loss out of the cavity to the total loss \cite{fabre_noise_1989, castelli_quantum_1994}.

It is essential to control the squeezing factor for applications such as quantum enhanced optical phase tracking and for the generation of Schr\"odinger cat states \cite{ourjoumtsev_generating_2006}. A large degree of squeezing in one field quadrature is inevitably concomitant with a large anti-squeezing in the orthogonal quadrature, which increases the phase estimation error \cite{yonezawa_quantum-enhanced_2012, caves_quantum-mechanical_1981}. As an example, Yonezawa \emph{et al.} reported a theoretically optimum degree of squeezing of around 7 dB for pure squeezed states, which reduces to $\sim$3.5 dB when unavoidable experimental imperfections such as losses and finite bandwidths are taken into account \cite{yonezawa_quantum-enhanced_2012}. The effect of anti-squeezing in the orthogonal quadrature is significant for phase tracking since it involves detecting large excursions in phase space from the mean value of the signal. This is in contrast to other sensing applications where one is measuring small fluctuations about a large mean value. A tunable squeezing factor is also required to generate optical Schr\"odinger kittens or cats depending on the size of the desired cat state \cite{ourjoumtsev_generating_2006, laghaout_amplification_2013}.

Conventional OPO cavities limit us to a fixed degree of squeezing, since they are usually made of discrete mirrors with a fixed reflectivity or using microresonators with a certain coupling. The coupling condition can in principle be modified by changing the mirrors, but this is a cumbersome process, especially for the monolithic cavities used to achieve strong squeezing \cite{vahlbruch_observation_2008, ast_high-bandwidth_2013}, where the mirror coatings are deposited directly on the nonlinear crystal. More recently, tapered fibers \cite{cai_observation_2000} and diamond prisms \cite{furst_quantum_2011, fortsch_versatile_2013} have been used to control the evanescent coupling to whispering gallery mode microresonators, but this requires precise experimental control of the coupling gap at the tens-of-nm level. Note that the degree of squeezing below the oscillation threshold can be changed by varying the pump power \cite{reynaud_quantum_1987, savage_squeezing_1987}. However, above threshold, the degree of squeezing in the intensity difference is independent of the pump power. Another way to tune the squeezing factor is to variably attenuate a highly squeezed state, but this method is inefficient since it requires a large power to generate the initial highly squeezed state.

In this article, we demonstrate continuous tuning of the degree of squeezing based on our previous demonstration of optical squeezing in a fully integrated silicon nitride OPO \cite{dutt_-chip_2015}. The Si$_3$N$_4$ OPO \cite{levy_cmos-compatible_2010} operates based on the third order nonlinear process of four-wave mixing (FWM), which generates quantum correlated signal and idler ``twin" beams when excited by a pump laser \cite{dutt_-chip_2015}. Specifically, the noise in the intensity difference between the twin beams is reduced below the shot noise level, and this reduction or squeezing factor, above the oscillation threshold, can be quantified as \cite{castelli_quantum_1994, dutt_-chip_2015, brambilla_nondegenerate_1991, chembo_quantum_2014},

\begin{equation}
S(\Omega) = 1 - \frac{\eta_c \eta_d}{1+\Omega^2 \tau_c^2}
\label{eq:squeezing}
\end{equation}
where $S(\Omega)$ is the intensity difference noise normalized to the shot noise level at a sideband frequency $\Omega$; $\eta_c$  is the coupling efficiency, i.e. the ratio of the coupling loss to the total (intrinsic + coupling) loss; $\eta_d$ is the detection efficiency, which takes into account the losses from the chip to the detector as well as the nonideal quantum efficiency of the detectors; and $\tau_c$ is the photon lifetime in the cavity. We reported 1.7 dB of intensity difference squeezing in our previous work using a single ring resonator \cite{dutt_-chip_2015}.

\begin{figure}[htbp]
\centering
\includegraphics[width=.6\linewidth]{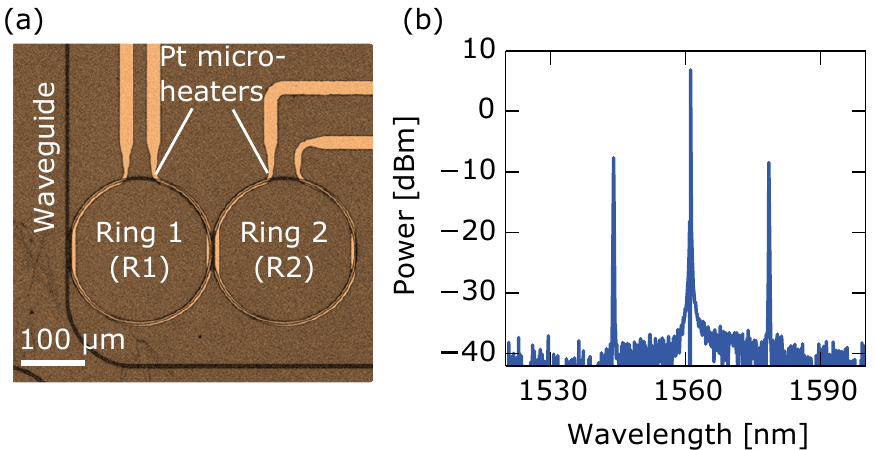}
\caption{(a) Optical microscope image of the coupled double-ring system used to tune the degree of squeezing. (b) Optical spectrum analyzer scan of the pump, signal and idler modes.}
\label{fig:device_spectrum}
\end{figure}

The tuning of the degree of squeezing is achieved here by externally controlling the coupling efficiency of a double-ring system. Our device consists of two coupled microring resonators, each with a radius of 115 \textmu m, as shown in Fig.~\ref{fig:device_spectrum}(a). One of the rings, R1, is coupled to a bus waveguide to launch light into the double ring system, while the second ring, R2, is coupled only to R1. By changing the offset between the resonance frequencies of R1 and R2, one can control the fraction of light in each ring and hence induce strong changes in the coupling condition of the double ring system to the bus waveguide. It should be noted that coupled cavity systems have been explored in various applications such as heralded single photon sources \cite{davanco_telecommunications-band_2012, kumar_spectrally_2013},  coupled-resonator-induced transparency \cite{smith_coupled-resonator-induced_2004}, four-wave mixing and frequency comb generation \cite{gentry_tunable_2014, wade_wavelength_2015, zeng_design_2014, xue_normal-dispersion_2015}. Here we utilize a coupled two-cavity system to tune the generated squeezing factor.

Our device uses microheaters integrated on double rings to thermally control the coupling efficiency $\eta_c$ between the rings and the bus waveguide and consequently tune the degree of squeezing.  The rings were fabricated on 950 nm thick Si$_3$N$_4$ films grown using low-pressure chemical vapor deposition (LPCVD), with 4 \textmu m of thermal oxide as the substrate. Trenches were defined in the wafer to mitigate stress-induced crack propagation in the nitride film, as described in \cite{luke_overcoming_2013}. The 1500 nm-wide waveguides were patterned with electron-beam lithography using ma-N 2405 resist on a JEOL 9500 system.  Both rings have 6 \textmu m wide platinum heaters which can be independently controlled to tune the resonance wavelength of the rings by the thermo-optic effect. To minimize losses caused by interaction of the optical modes with the platinum layer, the heaters and contacts were fabricated more than 2 \textmu m above the Si$_3$N$_4$ layer. By thermally tuning the resonance wavelength of the double ring system to the pump laser at 1561.1 nm, bright signal and idler beams were generated at 1544.1 nm and 1578.5 nm, as shown in Fig.~\ref{fig:device_spectrum}(b).

\begin{figure}[htbp]
\centering
\fbox{\includegraphics[width=.6\linewidth]{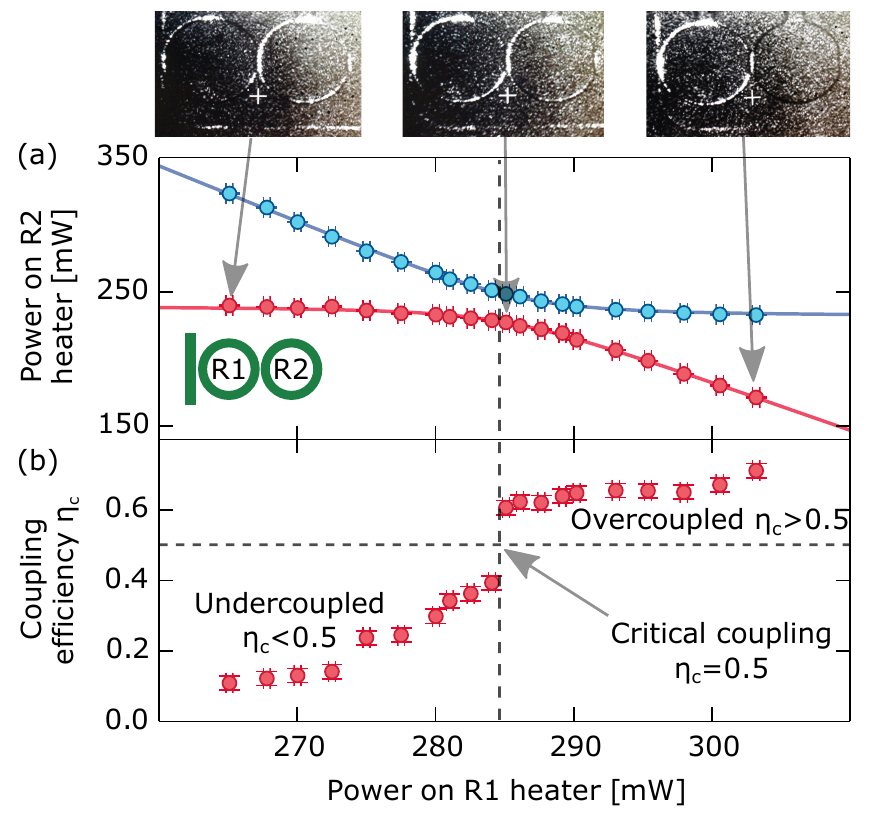}}
\caption{(a) Avoided crossing of the two-ring system, as the microheaters on the rings R1 and R2 are tuned to observe resonances at the fixed pump wavelength. The solid lines are fits to Eq.~\ref{eq:avoided_crossing}. From the top view IR images it is clear that light is initially localized in R2 and switches gradually to R1 as one tunes from one side of the avoided crossing to the other. (b) Coupling efficiency $\eta_c$ vs. the power on the R1 heater, for the lower red branch of the avoided crossing.}
\label{fig:avoided_crossing}
\end{figure}

The double ring system exhibits an avoided crossing when the heaters on R1 and R2 are tuned to match the resonance wavelengths of the system to the fixed laser wavelength, as shown in Fig.~\ref{fig:avoided_crossing}(a). For these measurements, the input power coupled into the waveguide is kept below 5 \textmu W to minimize thermal shift of the resonances due to the circulating optical power in the rings. From the top-view infrared (IR) camera images, we clearly see that the supermode of the double ring system undergoes a transition from being primarily localized in R2 to being localized in R1, for the lower red branch of the crossing. The solid lines in Fig.~\ref{fig:avoided_crossing}(a) are fits to the analytical expressions for the eigenfrequencies of the supermodes,

\begin{equation}
\omega_{\pm} = \frac{\omega_1 + \omega_2}{2} \pm \sqrt{\left(\frac{\omega_1-\omega_2}{2}\right)^2 + \kappa_{12}^2}
\label{eq:avoided_crossing}
\end{equation}
where $\omega_1$ and $\omega_2$ are the resonance frequencies of the two rings R1 and R2, in isolation, and $\kappa_{12}$ represents the coupling coefficient between the rings. For the remainder of this paper, we choose to focus on the lower red branch, although similar tunable coupling efficiencies were also observed for the upper branch. Note that avoided crossings have been reported between the transverse electric (TE) and transverse magnetic (TM) resonance wavelengths of a single ring resonator in the context of comb generation \cite{ramelow_strong_2014}. On the contrary, our design explicitly uses the fundamental TE modes of two coupled rings to generate tunable squeezing.

We observe a continuous change in the coupling efficiency $\eta_c$ between the double-ring system and the bus waveguide from 10\% to 70\% when tuning the heaters from one side of the avoided crossing to the other, as illustrated in Fig.~\ref{fig:avoided_crossing}(b). The coupling efficiency is determined from the normalized on-resonance transmission minimum: $\eta_c=(1\pm\sqrt{T_{min}})/2$, where the + and –- signs represent  overcoupled ($\eta_c>0.5$) and undercoupled ($\eta_c<0.5$) regimes respectively. The maximum $\eta_c$  is limited by the coupling coefficient of the bus waveguide to R1. This change in coupling can be intuitively explained by looking at the supermodes of the two-ring system. The mode on the blue-detuned side of the avoided crossing has a weaker overlap with the bus waveguide than the mode on the red-detuned side, and hence the coupling efficiency increases as one moves from the blue-detuned side of the avoided crossing to the red-detuned side.

\begin{figure}[htbp]
\centering
\fbox{\includegraphics[width=.6\linewidth]{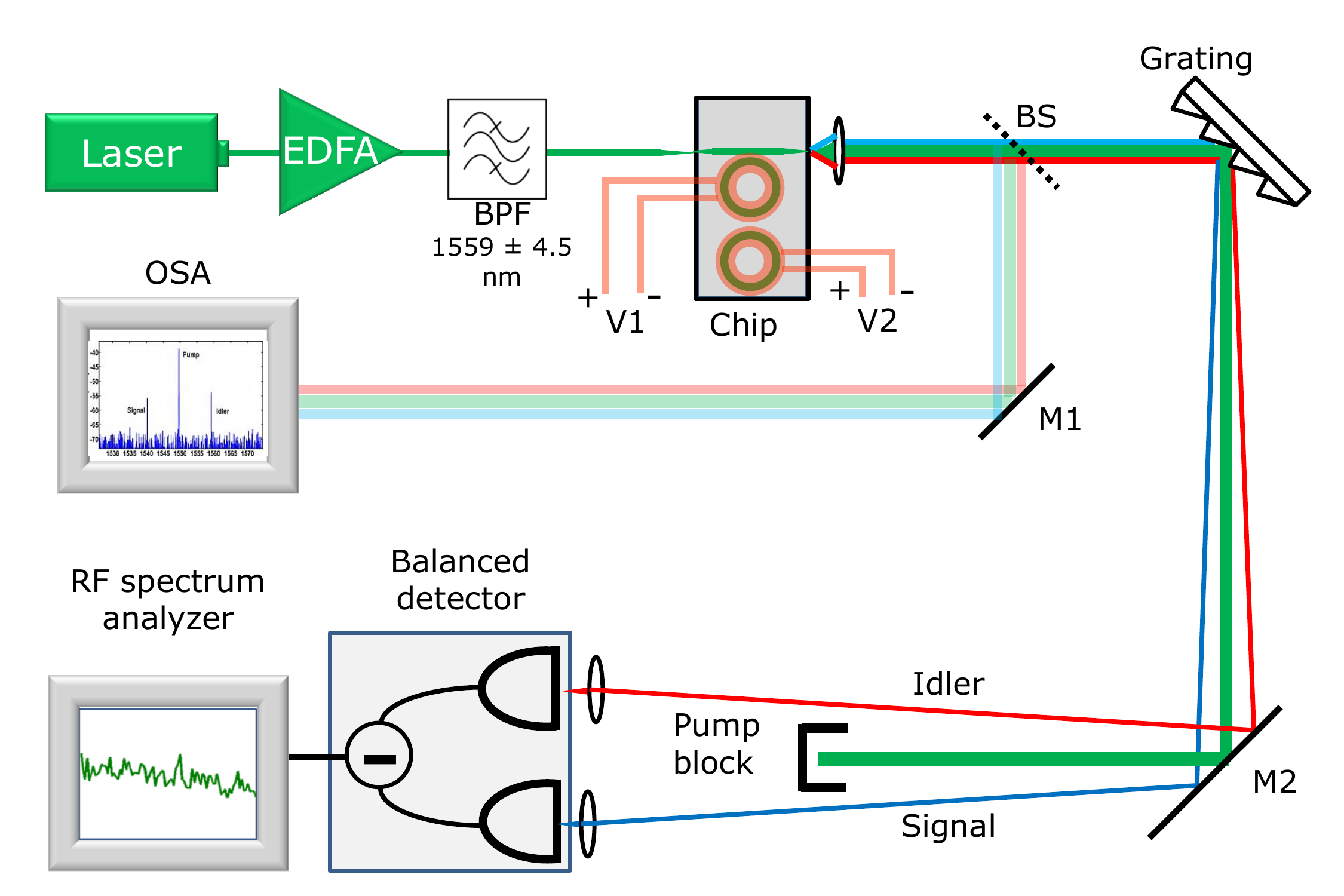}}
\caption{Experimental setup. EDFA: erbium doped fiber amplifier. BPF: bandpass filter. V1, V2: voltages applied to the microheaters on rings R1 and R2. BS: 1\% beam sampler, anti-reflection coated on one surface. M1, M2: mirrors. OSA: optical spectrum analyzer.}
\label{fig:exptsetup}
\end{figure}

The experimental setup used to generate and detect tunable squeezing is depicted in Fig.~\ref{fig:exptsetup}, and is similar to Ref. \cite{dutt_-chip_2015}. The rings are pumped by a low-noise single-frequency RIO Orion laser with a narrow linewidth $<3$ kHz \cite{numata_performance_2010}. Etched facet inverse tapers are used to efficiently mode match the lensed fiber to the input of the waveguides \cite{cardenas_high_2015}. The output of the chip is collected with an anti-reflection (AR) coated, high NA (NA = 0.56) aspheric lens to minimize losses to less than 0.7 dB. Electrical probes are used to apply voltages V1 and V2 to the contacts of the heaters on each of the two rings R1 and R2, respectively. A diffraction grating with an efficiency of 85\% spatially separates the pump, signal and idler beams. The signal and idler beams are directed to the two arms of a balanced detection system (Thorlabs PDB450C-AC), which consists of a pair of InGaAs photodiodes with a quantum efficiency of 80\%, followed by a low-noise transimpedance amplifier with a bandwidth of 4 MHz and a gain of $10^5$ V/A. The difference of the photocurrents is sent to an rf spectrum analyzer to measure the noise in the intensity difference between the two beams incident on the detectors. About 1\% of the output of the chip is sampled using a wedged beam sampler and sent to an optical spectrum analyzer (OSA) to monitor the onset of parametric oscillation.

\begin{figure}[htbp]
\centering
\fbox{\includegraphics[width=.6\linewidth]{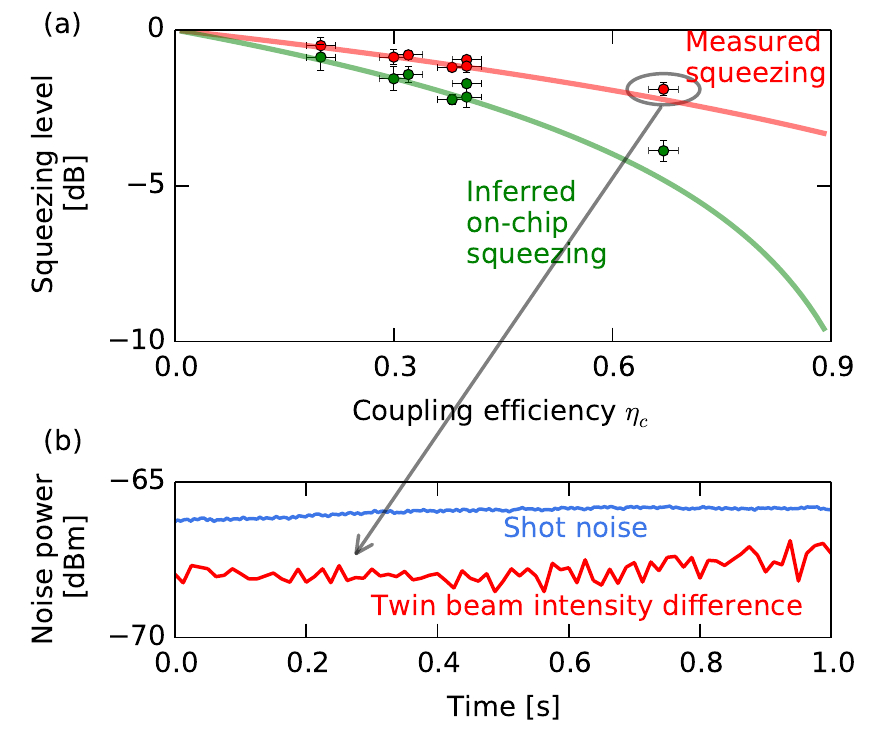}}
\caption{(a) The measured off chip squeezing (red) and inferred  on-chip (green) squeezing levels are plotted for various $\eta_c$. The data fit well to theoretical predictions based on the coupling and detection efficiencies (solid lines). Each data point in (a) is derived from time traces of the intensity difference noise, an example of which is shown in (b). (b) A time trace of the measured squeezing at a sideband frequency of $\Omega/2\pi = 3$ MHz (RBW: 30 kHz, VBW: 100 Hz). The dark noise is at -81 dBm.}
\label{fig:tunable_squeezing}
\end{figure}

We demonstrate a squeezing factor tunable from 0.5 dB in the strongly undercoupled regime to $\sim$2 dB in the overcoupled regime. In Fig.~\ref{fig:tunable_squeezing}(a), we plot the directly measured squeezing (red) in the intensity difference between the signal and idler beams, as a function of the coupling efficiency. Each data point in Fig.~\ref{fig:tunable_squeezing}(a) was obtained from time traces of the kind shown in Fig.~\ref{fig:tunable_squeezing}(b). The shot noise level was independently calibrated by splitting the pump on a 50:50 beam splitter and showed a linear dependence with optical power, as expected from theory. All measurements were taken at a sideband frequency of $\Omega/2\pi=3$ MHz, using a spectrum analyzer with a resolution bandwidth (RBW) of 30 kHz and a video bandwidth (VBW) of 100 Hz. By accounting for the overall detection efficiency of 60\%, we infer that the generated on-chip squeezing factor can be tuned from 0.9 $\pm$ 0.4 dB to 3.9 $\pm$ 0.3 dB. The measured and inferred squeezing levels match reasonably well with the predictions of Eq.~\ref{eq:squeezing} for $\eta_d = 60\%$ and $\eta_d = 100\%$ respectively.

The largest squeezing factor measured in our experiment was restricted by the maximum coupling efficiency of 70\% that can be achieved in the current device. By using devices with a smaller coupling gap between the bus and R1, this factor could be increased even further. On the other hand, the smallest squeezing level that could be reliably resolved was limited by the fluctuations in the time traces of the intensity difference noise. The minimum detectable squeezing can in principle be reduced by video averaging or by using a smaller resolution bandwidth, but this increases the data acquisition time and would require greater long-term stability in the operation of the OPO than was possible in the setup. 

In summary, we have introduced a technique to continuously change the degree of squeezing generated in an OPO by inducing strong changes in the coupling condition through integrated thermal tuning. This technique can pave the way for on-chip realization of quantum-enhanced sensing protocols where an optimal degree of squeezing is necessary for attaining the maximum allowed sensitivity.  The method demonstrated here can be extended to various on-chip platforms such as aluminum nitride, silica, diamond and silicon in which FWM oscillation has been reported \cite{jung_optical_2013, kippenberg_kerr-nonlinearity_2004, hausmann_diamond_2014, griffith_silicon-chip_2015}. Specifically, in aluminum nitride, the non-vanishing second order nonlinearity can enable fast electro-optic tuning of the coupling condition through the Pockels effect \cite{jung_electrical_2014}, as opposed to the slower thermal response of the microheaters used in the present work. More generally, the concept of externally tuning the effective coupling coefficient by utilizing supermodes of a multiple-cavity system finds applications in several areas of classical and quantum optics. For example, it can be used to tune the conversion efficiency and bandwidth of frequency comb generation \cite{miller_tunable_2015}. Thus, coupled resonators are a useful tool to control nonlinear processes and the resultant nonclassical states of the light field generated in them.

We gratefully acknowledge support from DARPA  under the Quiness program, and from AFOSR for awards \#BAA-AFOSR-2012-02 and \#FA9550-12-1-0377 supervised by Dr. Enrique Parra. P.N. acknowledges partial support from Fundação de Amparo à Pesquisa do Estado de São Paulo (FAPESP grant \#2011/12140-6). This work was performed in part at the Cornell NanoScale Facility, a member of the National Nanotechnology Infrastructure Network, supported by the National Science Foundation (Grant ECCS-0335765). This work made use of the Cornell Center for Materials Research Shared Facilities which are supported through the NSF MRSEC program (DMR-1120296).

We thank Carl Poitras for productive discussions and Sasikanth Manipatruni for performing initial experiments on squeezing.

\bigskip

\end{document}